\documentclass[aps,twocolumn,showpacs,byrevtex,prl,reprint,nofootinbib]{revtex4-1}
\usepackage{xspace}

\usepackage{epsfig}
\usepackage{dcolumn}
\usepackage{bm}
\usepackage{color}
\usepackage{graphicx}
\usepackage{subfigure}
\usepackage{pstricks}
\usepackage{pst-node}
\usepackage{rotating}
\usepackage{times}
\usepackage[english]{babel}
\addto{\captionsenglish}{%

}
\usepackage[colorlinks,linkcolor=blue,citecolor=blue]{hyperref}
\usepackage{lineno}
\newcommand{\XXB}{\Xi^{-}\bar\Xi^{+}}

\newcommand{\EE}{e^+e^-}

\newcommand{\psp}{\psi(3686)}
\newcommand{\jpsi}{J/\psi}
\newcommand{\ar}{\rightarrow}
\newcommand{\llb}{\Lambda\bar{\Lambda}}

\newcommand{\bfg}{\begin{figure}}
\newcommand{\efg}{\end{figure}}
\newcommand{\bitm}{\begin{itemize}}
\newcommand{\eitm}{\end{itemize}}
\newcommand{\bnum}{\begin{enumerate}}
\newcommand{\enum}{\end{enumerate}}
\newcommand{\btbl}{\begin{table*}}
\newcommand{\etbl}{\end{table*}}
\newcommand{\btbu}{\begin{tabular}}
\newcommand{\etbu}{\end{tabular}}
\newcommand{\bcl}{\begin{center}}
\newcommand{\ecl}{\end{center}}
\newcommand{\bbt}{\bibitem}
\newcommand{\beq}{\begin{equation}}
\newcommand{\eeq}{\end{equation}}
\newcommand{\beqr}{\begin{eqnarray}}
\newcommand{\eeqr}{\end{eqnarray}}

\begin{document}
\normalsize
\parskip=5pt plus 1pt minus 1pt
\title{\boldmath Measurement of the cross section for $\EE\ar\XXB$ and observation of an excited $\Xi$ baryon}
\author{
M.~Ablikim$^{1}$, M.~N.~Achasov$^{10,d}$, P.~Adlarson$^{59}$, S. ~Ahmed$^{15}$, M.~Albrecht$^{4}$, M.~Alekseev$^{58A,58C}$, A.~Amoroso$^{58A,58C}$, F.~F.~An$^{1}$, Q.~An$^{55,43}$, Y.~Bai$^{42}$
% ...
\\
\vspace{0.2cm}
(BESIII Collaboration)\\
\vspace{0.2cm} {\it
$^{1}$ Institute of High Energy Physics, Beijing 100049, People's Republic of China\\
% ...
$^{4}$ Bochum Ruhr-University, D-44780 Bochum, Germany\\
% ...
$^{10}$ G.I. Budker Institute of Nuclear Physics SB RAS (BINP), Novosibirsk 630090, Russia\\
% ...
$^{15}$ Helmholtz Institute Mainz, Johann-Joachim-Becher-Weg 45, D-55099 Mainz, Germany\\
% ...
$^{42}$ Southeast University, Nanjing 211100, People's Republic of China\\
$^{43}$ State Key Laboratory of Particle Detection and Electronics, Beijing 100049, Hefei 230026, People's Republic of China\\
% ...
$^{55}$ University of Science and Technology of China, Hefei 230026, People's Republic of China\\
% ...
$^{58}$ (A)University of Turin, I-10125, Turin, Italy; (B)University of Eastern Piedmont, I-15121, Alessandria, Italy; (C)INFN, I-10125, Turin, Italy\\
$^{59}$ Uppsala University, Box 516, SE-75120 Uppsala, Sweden\\
% ...
$^{d}$ Also at the Novosibirsk State University, Novosibirsk, 630090, Russia\\
% ...
}
}

\date{\today}

\begin{abstract}
  Using a total of 11.0 fb$^{-1}$ of $\EE$ collision data with
  center-of-mass energies between 4.009 GeV and 4.6 GeV and collected
  with the BESIII detector at BEPCII, we measure fifteen exclusive
  cross sections and effective form factors for the process
  $\EE\ar\XXB$ by means of a single baryon-tag method.  After
  performing a fit to the dressed cross section of $\EE\ar\XXB$, no
  significant $\psi(4230)$ or $\psi(4260)$ resonance is observed in
  the $\XXB$ final states, and upper limits at the 90\% confidence
  level on $\Gamma_{ee}\mathcal{B}$ for the processes
  $\psi(4230)$/$\psi(4260)\ar\XXB$ are determined.  In addition, an
  excited $\Xi$ baryon at 1820 MeV/$c^{2}$ is observed with a
  statistical significance of  6.2 $\sim$ 6.5$\sigma$ by including the systematic uncertainty, and the mass and width are
  measured to be $M = (1825.5 \pm 4.7 \pm 4.7)$~MeV/$c^{2}$ and
$\Gamma = (17.0 \pm 15.0 \pm 7.9)$~MeV, which confirms the existence of the $J^{P}=\frac{3}{2}^{-}$ state $\Xi(1820)$.
\end{abstract}
\maketitle

In the last decade, a series of charmonium-like states have been
observed at $\EE$ colliders.  The study of the production of
charmonium-like states with the quantum number $J^{PC} = 1^{--}$ above
open charm threshold in $\EE$ annihilations and their subsequent
two-body hadronic decays provides a test for QCD
calculations~\cite{Farrar,Briceno}.  According to potential models,
there are five vector charmonium states between the $1D$ state
$(\psi(3773))$ and 4.7 GeV/$c^{2}$, namely, the $3S$, $2D$, $4S$,
$3D$, and $5S$ states~\cite{Farrar}. From experimental studies,
besides the three well-established structures observed in the
inclusive hadronic cross section~\cite{PDG2016}, \emph{i.e.}, $\psi(4040)$,
$\psi(4160)$, and $\psi(4415)$, five new states, \emph{i.e.}, $\psi(4230)$,
$\psi(4260)$, $\psi(4360)$, $\psi(4634)$, and $\psi(4660)$ have been
reported in initial state radiation (ISR) processes, \emph{i.e.},
$\EE\ar\gamma_{ISR}\pi^{+}\pi^{-}\jpsi$ or
$\EE\ar\gamma_{ISR}\pi^{+}\pi^{-}\psp$ at the BABAR~\cite{BABAR01} and Belle~\cite{BELLE01}, or in direct production processes at
the CLEO~\cite{CLEO} and BESIII experiments~\cite{BESIIIAB}.
Surprisingly, up to now, no evidence for baryon anti-baryon pairs
above open charm production associated with these states has been
found except for the $\psi(4634)$ resonance observed in
$\Lambda^{+}_{c}\bar\Lambda_{c}^{-}$~\cite{LCLCbar}.  Although the
BESIII Collaboration previously performed a search for baryonic decays
of $\psi(4040)$~\cite{Ablikim:2013pgf}, including $\XXB$ final states
based on a full reconstruction method, no candidates were observed.
The overpopulation of structures in this mass region and the mismatch of
the properties between the potential model predictions and experimental
measurements make them good candidates for exotic states. Various
scenarios, which interpret one or some of them as hybrid states,
tetraquark states, or molecular states~\cite{QCD}, have been proposed.

The electromagnetic structure of hadrons, parameterized in
terms of electromagnetic form factors (EMFFs)~\cite{ABC}, provides a key to
understanding QCD effects in bound states. While the nucleon has been
studied rigorously for more than sixty years, new techniques and
the availability of data with larger statistics from modern facilities
have given rise to a renewed interest in the field, \emph{i.e.}, the proton
radius puzzle~\cite{Nature}.
The access to hyperon structure by EMFFs provides an extra  dimension
that inspires  measurements of exclusive cross
sections and EMFFs for baryon anti-baryon pairs above open charm
threshold.

The constituent quark model has been very successful in describing the
ground state of the flavor SU(3) octet and decuplet
baryons~\cite{PDG2016,Chao:1980em}. However, some observed excited
states do not agree well with the theoretical prediction. It is thus
important to study such unusual states, both to probe the limitation
of the quark models and to spot unrevealed aspects of the QCD
description of the structure of hadron resonances. Intriguingly, the
$\Xi$ resonances with strangeness $S = -2$ may provide important
information on the latter aspect.  Although, there has been
significant progress in the experimental studies of charmed baryons by
the BaBar~\cite{Aubert:2006sp}, LHCb~\cite{Aaij:2017nav}, and
Belle~\cite{Yelton:2017qxg,Li:2018fmq} collaborations, doubly-charm
baryons by the LHCb collaboration~\cite{Aaij:2017ueg}, doubly-strange
baryons by the Belle collaboration~\cite{Yelton:2018mag}, the studies of excited $\Xi$ states are still
sparse~\cite{PDG2016}.  Neither the first radial excitation with 
spin-parity of $J^{P} =\frac{1}{2}^{+}$ nor a first orbital excitation with
$J^{P} =\frac{1}{2}^{-}$ have been identified.  Determination of the resonance
parameters of the first excited state is a vital test of our
understanding of the structure of $\Xi$ resonances, where one of
candidates for the first excited state is $\Xi(1690)$ with a
three-star rating on a four-star scale~\cite{PDG2016}, the second one is $\Xi(1620)$ with
one-star rating, and another excited state is $\Xi(1820)$ with a
three-star rating~\cite{PDG2016}, for which the spin was previously
determined to be $J = \frac{3}{2}$~\cite{Teodoro:1978bu}, and
subsequently the parity was determined to be negative and the spin-parity
confirmed to be $J^{P} = \frac{3}{2}^{-}$ by another
experiment~\cite{Biagi:1986vs}.

In this Letter, we present a measurement of the Born cross section
and the effective form factors (EFF)~\cite{ABC} for the process $\EE\ar\XXB$, an
estimation of the upper limit on
$\Gamma_{ee}\mathcal{B}$($\psi(4230)$/$\psi(4260)\ar\XXB$) at the 90\%
confidence level (CL), and the observation of an excited $\Xi$ baryon
at 1820 MeV/$c^{2}$.  The dataset used in this analysis
corresponds to a total of 11.0 fb$^{-1}$ of $\EE$ collision data~\cite{ABC}
collected at center-of-mass (CM) energies from 4.009 GeV to 4.6 GeV
with the BESIII detector~\cite{Wang:2007tv} at BEPCII~\cite{BESIII} .

The selection of $\EE\ar\XXB$ events with a full reconstruction method
has low-reconstruction efficiency.  Here, to achieve higher efficiency, a
single baryon $\Xi^{-}$ tag technique is employed, \emph{i.e.}, only one
$\Xi^{-}$ baryon is reconstructed by the $\pi^{-}\Lambda$ decay mode with
$\Lambda\ar p\pi^{-}$ , and the anti-baryon $\bar\Xi^{+}$ is extracted
from the recoil side (unless otherwise noted, the charge-conjugate
state of the $\Xi^{-}$ mode is included by default below).  To
determine the detection efficiency for the decay $\EE\ar\XXB$,
100k simulated events are generated for each of 15 energy points in the range of 4.009 to 4.6 GeV according
to phase space using the \textsc{kkmc} generator~\cite{kkmc}, which
includes the ISR effect.  The $\Xi^{-}$ is simulated in its decay to
the $\pi^{-}\Lambda$ mode with the subsequent decay $\Lambda\to
p\pi^{-}$ via \textsc{evtgen}~\cite{evt2}, and the anti-baryons are
allowed to decay inclusively.
The response of the BESIII detector is modeled with Monte Carlo (MC) simulations
using a framework based on \textsc{geant}{\footnotesize 4}~\cite{geant4}.
 Large simulated samples of generic $\EE \to
\text{hadrons}$ events (`inclusive MC') are used to estimate
background conditions.

Charged tracks are required to be reconstructed in the main drift
chamber (MDC) with good helical fits and within the angular coverage
of the MDC: $|\cos\theta|<0.93$, where $\theta$ is the polar angle
with respect to the $e^{+}$ beam direction.  Information from the
specific energy deposition ($dE/dx$) measured in the MDC combined with
the time-of-flight (TOF) is used to form particle identification (PID)
confidence levels for the hypotheses of a pion, kaon, and proton.
Each track is assigned to the particle type with the highest
CL. Events with at least two negatively charged pions and one proton
are kept for further analysis.

To reconstruct $\Lambda$ candidates, a secondary vertex fit is
applied to all $p\pi^{-}$ combinations; the ones characterized by
$\chi^{2} < 500$ with 3 degrees of freedom are kept for further analysis. The $p\pi^{-}$
invariant mass is required to be within 5 MeV/$c^{2}$ of the nominal
$\Lambda$ mass, determined by optimizing the figure of merit
$\frac{S}{\sqrt{S + B}}$ based on the MC simulation, where $S$ is the
number of signal MC events and $B$ is the number of the background events expected from simulation.
To further suppress background from non-$\Lambda$ events, the $\Lambda$ decay
length is required to be greater than zero, where
negative decay lengths are caused by the limited detector
resolution.  

The $\Xi^{-}$ candidates are reconstructed with a similar strategy
using a secondary vertex fit, and the candidate with the minimum value
of $|M_{\pi^{-}\Lambda}-m_{\Xi^{-}}|$ from all $\pi^{-}\Lambda$
combinations is selected, where $M_{\pi^{-}\Lambda}$ is the invariant
mass of the $\pi^{-}\Lambda$ pair, and $m_{\Xi^{-}}$ is the nominal
mass of $\Xi^{-}$ from the PDG~\cite{PDG2016}. Further
$M_{\pi^{-}\Lambda}$ is required to be within 10 MeV/$c^{2}$ of the
nominal $\Xi^{-}$ mass, and the $\Xi^{-}$ decay length $L_{\Xi^{-}}$ (cm)
is required to be greater than zero.

\begin{figure}[!htbp]
\includegraphics[width=0.3\textwidth]{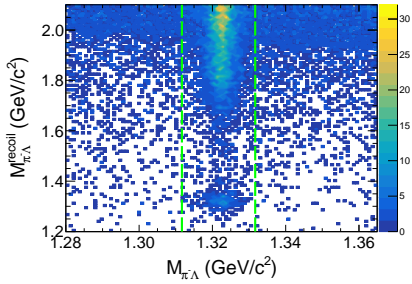}
\caption{Distribution of $M_{\pi^{-}\Lambda}$ versus $M^{\rm
    recoil}_{\pi^{-}\Lambda}$ for sum of 15 energy points. The dashed
  lines denote the $\Xi^{-}$ signal region.}
\label{scatterplot}
\end{figure}

To obtain the anti-baryon candidates $\bar\Xi^{+}$, we use the
distribution of mass recoiling against the selected $\pi^{-}\Lambda$
system,
\begin{equation}
  M^{\rm recoil}_{\pi^{-}\Lambda} = \sqrt{(\sqrt{s}-E_{\pi^{-}\Lambda})^{2} - |\vec{p}_{\pi^{-}\Lambda}|^{2}},
\end{equation}
where $E_{\pi^{-}\Lambda}$ and $\vec{p}_{\pi^{-}\Lambda}$ are the
energy and momentum of the selected $\pi^{-}\Lambda$ candidate in the
CM system, and $\sqrt{s}$ is the CM energy.  
Figure~\ref{scatterplot} shows the distribution of
$M_{\pi^{-}\Lambda}$ versus $M^{\rm recoil}_{\pi^{-}\Lambda}$ for all 15 considered energy points.

\begin{figure}[!htbp]
\includegraphics[width=0.4\textwidth]{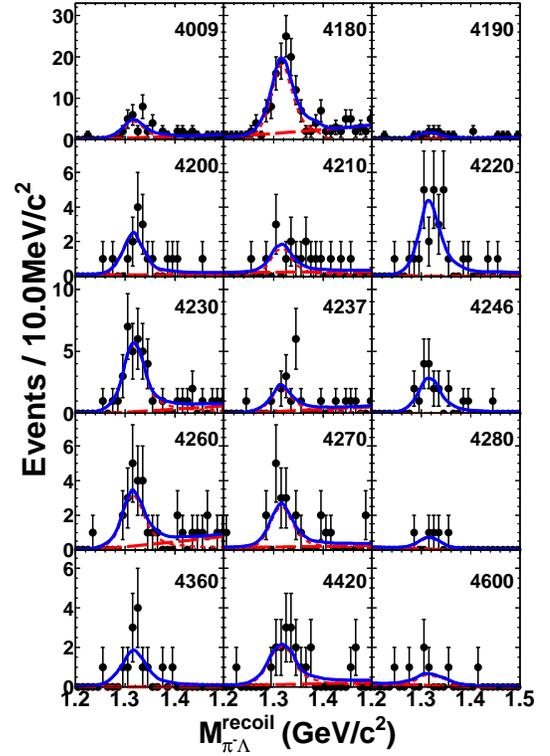}
\caption{Fit to the recoil mass spectra of $\pi^{-}\Lambda$ at each energy
  point in units of MeV/$c^2$.  Dots with error bars are data, the blue solid
  lines show the fit result, the red short-dashed lines are for
  signal, and the red long-dashed ones are for the smooth background.}
\label{scatter_recoil_fitting}
\end{figure}

The signal yields for the decay $\EE\ar\XXB$ at each energy point are
determined by performing an extended maximum likelihood fit to the
$M^{\rm recoil}_{\pi^{-}\Lambda}$ spectrum in the range from 1.2
GeV/$c^{2}$ to 1.5 GeV/$c^{2}$. In the fit, the signal shape for the
decay $\EE\ar\XXB$ at each energy point is represented by the
simulated MC shape.  After applying the same event selection as the
data on the inclusive MC samples at each CM energy, it is found that
few background events remain at each energy point coming from
$\EE\ar\pi^{+}\pi^{-}\jpsi$, $\jpsi\ar\llb$ events, and they are
distributed smoothly in the region of interest and can be
described by a second-order polynomial function.
Figure~\ref{scatter_recoil_fitting} shows the $M^{\rm
  recoil}_{\pi^{-}\Lambda}$ distributions for the decay $\EE\ar\XXB$
at each energy point.

The Born cross section for $\EE\ar\XXB$ is calculated by
\begin{equation}
\sigma^{B}(s) =\frac{N_{\rm obs}}{2{\cal{L}}(1 + \delta)\frac{1}{|1 - \Pi|^2}\epsilon{\cal B}(\Xi^{-}\ar\pi^{-}\Lambda){\cal B}(\Lambda\ar p\pi^{-})},
\end{equation}
where $N_{\rm obs}$ is the number of the observed signal events,
${\cal{L}}$ is the integrated luminosity related to the CM energy, $(1
+ \delta)$ is the ISR correction factor~\cite{Jadach:2000ir}, $\frac{1}{|1 - \Pi|^2}$ is the vacuum
polarization correction factor~\cite{Actis:2010gg}, $\epsilon$ is the detection
efficiency, and $ {\cal B}(\Xi\ar\pi^{-}\Lambda)$ and ${\cal
  B}(\Lambda\ar p\pi^{-})$ are the branching fractions taken from the
PDG~\cite{PDG2016}.  The ISR correction factor is obtained using the
QED calculation as described in Ref.~\cite{Kuraev:1985hb} and taking
the formula used to fit the cross section measured in this analysis
parameterized after two iterations as input.  The measured cross
sections and EFFs are shown in Fig.~\ref{Cross_SC_com} and summarized in the Supplemental Material~\cite{ABC}.  The Supplemental Material also
contains the details of the cross section and EFF
calculations.

A maximum likelihood method is used to fit the dressed cross section
$\sigma^{\rm dressed} = \ \sigma^{B}/|1 - \Pi|^{2}$ for the process
$\EE\ar\XXB$ parameterized as the coherent sum of a power-law function
plus a Breit-Wigner (BW) function for $\psi(4230)$ or $\psi(4260)$,
\begin{equation}
    \sigma^{\rm dressed}(\sqrt{s})= |c_{0}\frac{\sqrt{P(\sqrt{s})}}{s^{n}} + e^{i\phi}{\rm BW}(\sqrt{s})\sqrt{\frac{P(\sqrt{s})}{P(M)}}|^{2},
\end{equation}
where the mass, $M$, and total width, $\Gamma$, are fixed to the
$\psi(4230)/\psi(4260)$ resonance with PDG values~\cite{PDG2016},
$\phi$ is the relative phase between the BW function,
\begin{equation}
    {\rm BW}(\sqrt{s}) =\frac{\sqrt{12\pi\Gamma_{ee}{\cal{B}}\Gamma}}{s-M^{2}+iM\Gamma},
\end{equation}
and power function, $n$ is a free fit parameter, and $P(\sqrt{s})$ is the two-body phase space factor. The
$\psi(4230)$ and $\psi(4260)\ar\XXB$ processes are found to be not significant. Therefore,
upper limits on the products of the two-electron partial width and the
branching fractions of $\psi(4230)$ and $\psi(4260)\ar\XXB$
$(\Gamma_{ee}{\cal{B}})$ at the 90\% credible level are estimated using a Bayesian
approach~\cite{ZHUYS} to be $\Gamma_{ee}\mathcal{B}_{\psi{(4230)}} <
0.33 \times 10^{-3}$~eV and $\Gamma_{ee}\mathcal{B}_{\psi{(4260)}} <
0.27 \times 10^{-3}$~eV taking into account the systematic uncertainty
described later. Here the masses and widths of $\psi(4230)$ and
$\psi(4260)$ are changed by all combinations of $\pm 1 \sigma$, and the
estimation of the upper limits repeated. The largest ones are taken as
the final results. Figure~\ref{Cross_SC_com} shows the fit to the
dressed cross section assuming the $\psi(4230)$ or the $\psi(4260)$
resonance and without resonance assumption. Including systematic uncertainties, the significance for both resonances is calculated to be $\sim$2.7$\sigma$.

The EFF for $\EE\ar\XXB$ is calculated by the formula~\cite{ABC},
\begin{equation}
|G_{\rm eff}(s)|  = \sqrt{\frac{3s\sigma^{B}}{4\pi\alpha^2C\beta(1+\frac{2m^{2}_{\Xi^{-}}}{s})}},
\end{equation}
where 
$\alpha$ is the fine structure constant, $m_{\Xi^{-}}$ is the mass of
$\Xi^{-}$, the variable $\beta =\sqrt{1-\frac{1}{\tau}}$ is the
velocity, $\tau = \frac{s}{4m^{2}_{\Xi^{-}}}$, and the Coulomb
correction factor $C$~\cite{Baldini} parameterizes the
electro-magnetic interaction between the outgoing baryon and
anti-baryon. For neutral baryons, the Coulomb factor is unity,
while  for point-like charged fermions $ C =
\frac{\pi\alpha}{\beta}\cdot\frac{\sqrt{1-\beta^2}}{1-e^{-\frac{\pi\alpha}{\beta}}}
$~\cite{Sommerfeld,
 Tzara, Sakharov}. Figure~\ref{Cross_SC_com} shows the measured EFFs of
$\EE\ar\XXB$.
\begin{figure}[!htbp]
\includegraphics[width=0.25\textwidth]{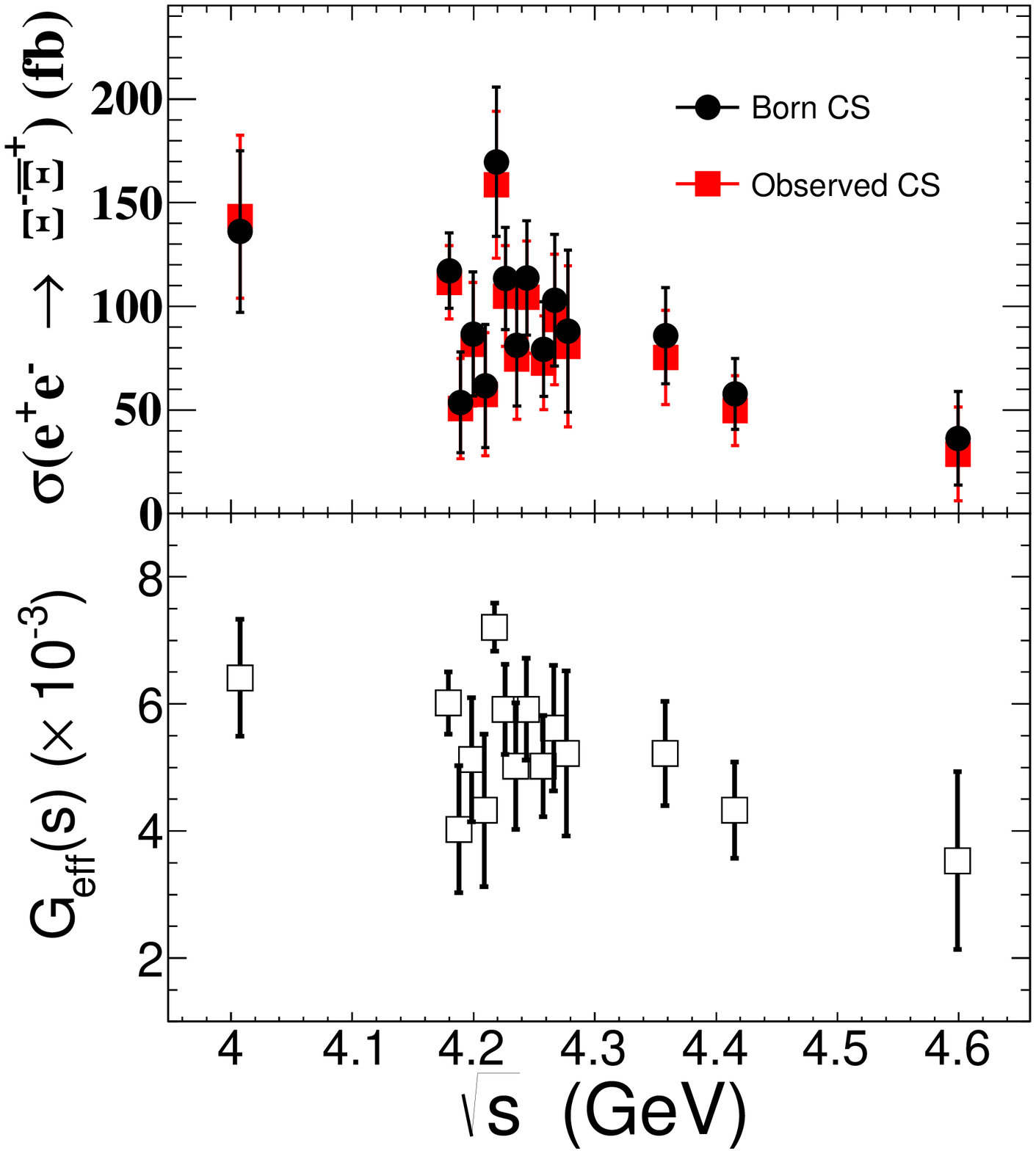}\\
\includegraphics[width=0.15\textwidth]{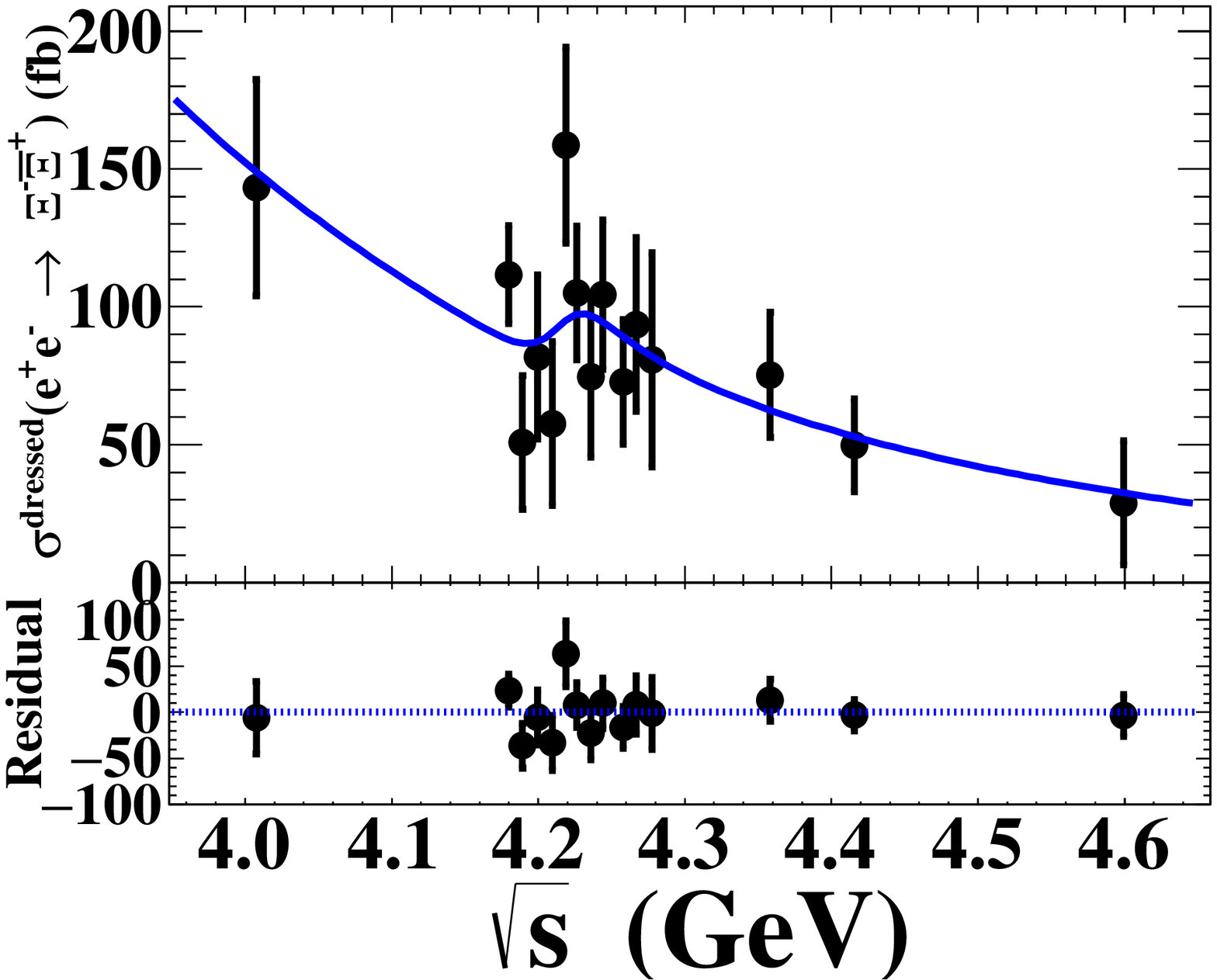}
\includegraphics[width=0.15\textwidth]{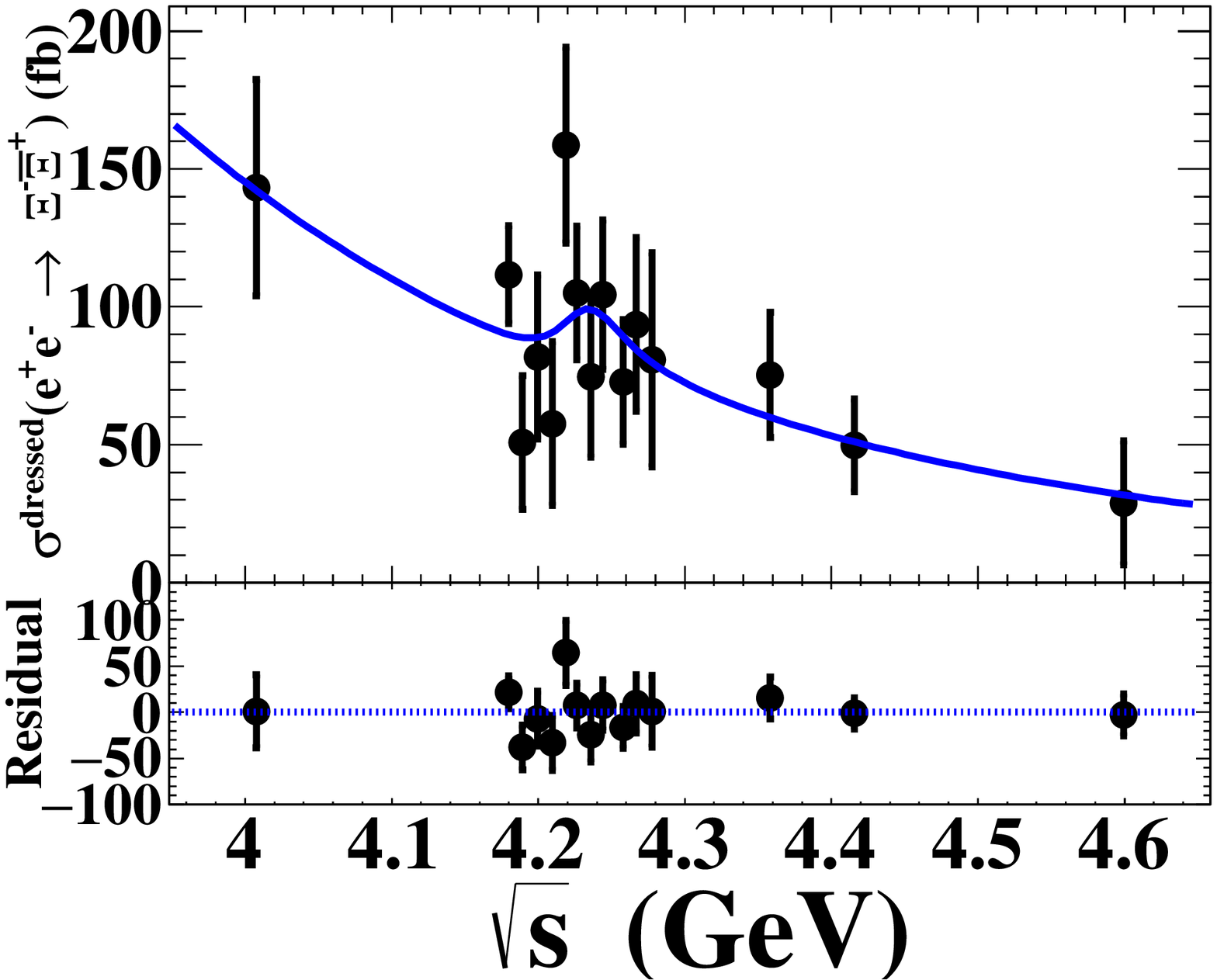}
\includegraphics[width=0.15\textwidth]{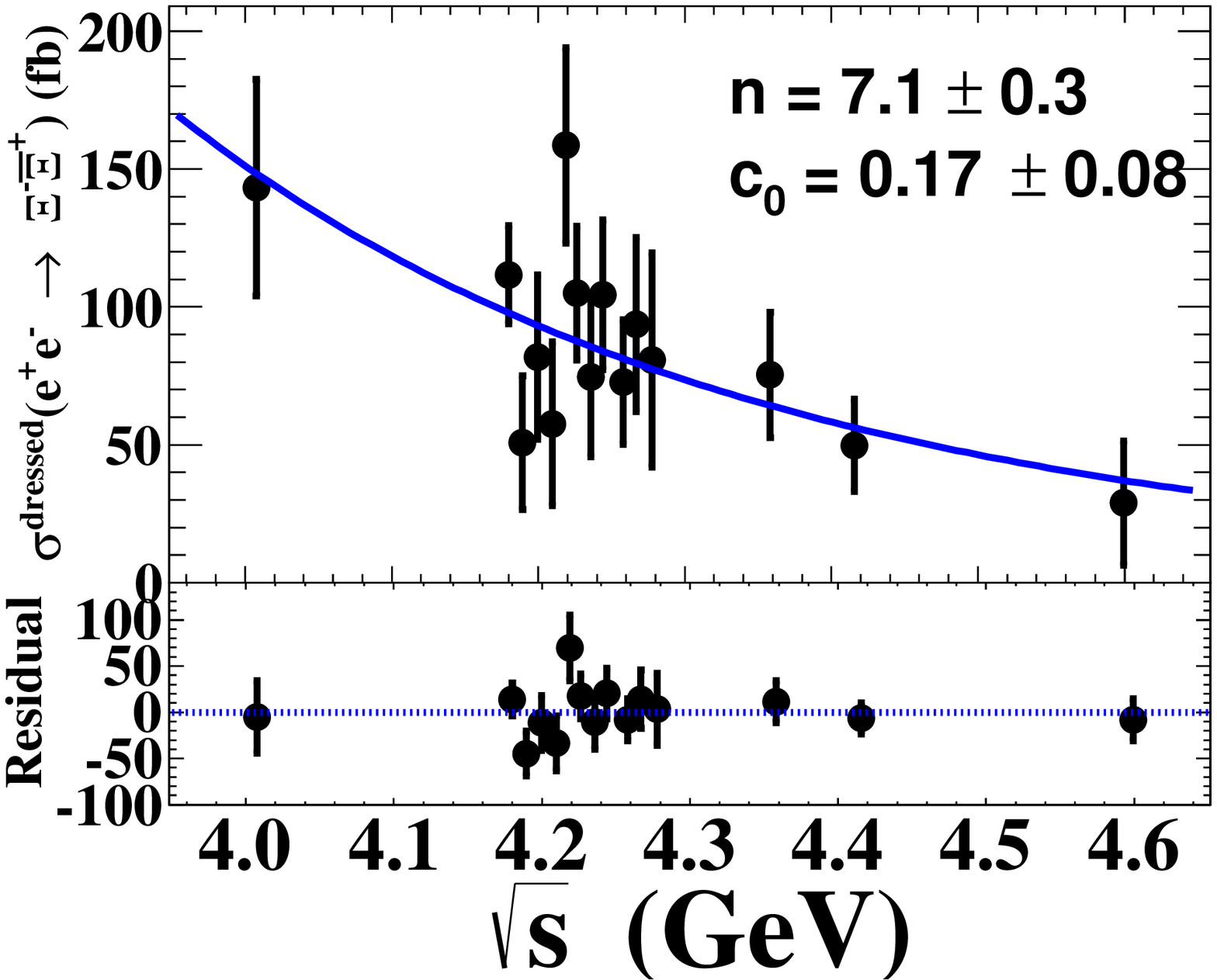}
\caption{ Top:  cross section (points with error bars) and EFF (open
  boxes with error bars).
 Bottom:  fits to the dressed cross sections at CM energies from 4.009 to 4.6 GeV with the assumptions
 of a power-law function plus a $\psi(4230)$ resonance function (Left) or a $\psi(4260)$ resonance function (Middle), and without resonance assumption (Right)
 where the dots with error bars are the dressed cross sections and the
 solid lines show the fit results.}
\label{Cross_SC_com}
\end{figure}

\begin{figure}[!htbp]
\includegraphics[width=0.32\textwidth]{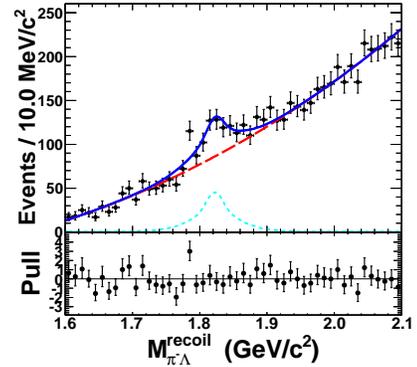}
\caption{Fit to the recoil mass of $\pi\Lambda$ of the combined data
  of the 15 energy points in the range from 1.6 to 2.1 GeV/$c^{2}$.  Dots
  with error bars are data, the blue solid line shows the fit result,
  the short-dashed line is for signal, and the long-dashed one is for the
  background.}
\label{scatterplot_1820}
\end{figure}

Based on the selected data for the sum of 15 energy points, an excited
$\Xi$ baryon is observed in the $M^{\rm recoil}_{\pi^{-}\Lambda}$
range from 1.6~GeV/$c^{2}$ to 2.1~GeV/$c^{2}$.
Figure~\ref{scatterplot_1820} shows a fit to the recoil mass spectrum of
$\pi^{-}\Lambda$, where the signal is described by a BW function convolved
with a double Gaussian function, and the background is described by a
$2^{nd}$ order Chebyshev polynomial, where the resolution width of
Gaussian function is fixed according to the MC simulation. The number
of signal events is $288^{+125}_{-85} $, and the mass and width are measured
to be $ M = (1825.5 \pm 4.7)$ MeV/$c^{2}$ and $ \Gamma= (17.0 \pm
15.0)$ MeV, where the uncertainties are statistical only.  The
statistical significance of the 1820 MeV/$c^{2}$  resonance is estimated to be
6.2 $\sim$ 6.5 standard deviation with including the systematic uncertainty.

Systematic uncertainties on the measurements of the cross section
originate from the luminosity measurement, branching fractions of
$\Xi^{-}\ar\pi^{-}\Lambda$ and $\Lambda\ar p\pi^{-}$, detection
efficiency, ISR correction factor, line-shape structure, angular
distribution, and the fit procedure. The uncertainty due to the vacuum polarization
is negligible.  The integrated luminosity is measured with a precision of 1.0\%~\cite{Ablikim:2015nan}.  The branching fraction
uncertainties for $\Xi^{-}\ar\pi^{-}\Lambda$ and $\Lambda\ar p\pi^{-}$
are 0.1\% and 0.8\% from the PDG~\cite{PDG2016}.  The systematic
sources of the uncertainty for the detection efficiency include the
$\Xi^{-}$ reconstruction, the mass windows of $\Xi^{-}/\Lambda$, and
the decay lengths of $\Xi^{-}/\Lambda$. The $\Xi^{-}$ reconstruction
is studied using the same method as described in
Ref.~\cite{Ablikim:2016iym}, and an uncertainty of 6.6\% is found. The
mass windows of $\Xi^{-}$ and $\Lambda$ are studied by varying the
nominal requirements by 5.0 MeV/$c^{2}$, which yield uncertainties of
0.7\% and 3.2\%, respectively. The decay lengths of $\Lambda$ and $\Xi^{-}$ are
studied with and without the nominal requirements, and the
uncertainties are estimated to be 1.5\% and 1.7\%, respectively.  For
the ISR correction factor, we iterate the cross section measurement
until $(1+\delta)\epsilon$ converges as described in
Ref.~\cite{Ablikim:2016qzw}. The change due to the different criteria
for convergence is taken as the systematic uncertainty. The uncertainty
due to the line-shape structure is estimated to be 4.8\% with the
assumption of $\psi(4230)/\psi(4260)\ar\XXB$. The uncertainty due to
the angular distribution is estimated to be 4.0\% by weighting the
$\cos\theta_{\Xi}$ difference for each bin between the data and the
phase space MC model, where the $\theta_{\Xi}$ is the angle between
$\Xi$ and the beam directions in the $\EE$ CM
system~\cite{Ablikim:2016iym}.  The systematic sources of the
uncertainty in the fit of the $M^{\rm recoil}_{\pi^{-}\Lambda}$
spectrum include the fitting range, the polynomial shape, the mass
resolution, the signal shape the mass windows of $\Xi^{-}/\Lambda$,
and the decay lengths of $\Xi^{-}/\Lambda$. The uncertainty due to the
fit range is estimated to be 3.3\% by varying the mass range by
$\pm50$~MeV/$c^{2}$. The uncertainty due to the polynomial function is
estimated to be 3.3\% by alternative fits with a third- or a
first-order polynomial function.  The mass resolution is studied by
varying the nominal signal shape convolved with a Gaussian function,
and the yield difference is taken as a systematic uncertainty, which
is 4.0\%.  The effect due to the signal shape is studied by varying
the resolution in the convolution of the Breit-Wigner with a Gaussian
function. This gives an uncertainty of 3.2\%.  The effect of the
MC statistics on the used signal shape is studied by using an MC
sample with only 10\% of the events compared to the nominal fit,
and the uncertainty is 0.5\%.  Assuming all sources to be independent,
the total systematic uncertainty on the cross section measurement for
$\EE\ar\XXB$ is determined to be 12.7\% by the quadratic sum of these
sources.

Systematic uncertainties on the measurements of the mass and width for
the excited $\Xi$ state mainly originate from the fit range, the
background shape, the mass resolution and the signal shape.  The
fit range, the background, and signal shapes are studied with the same
method as above with mass uncertainties of 1.5,
1.3, and 1.9~MeV/$c^2$, and width uncertainties of 5.6,
3.4, and 4.5~MeV, respectively.  The mass uncertainty due to the
mass resolution is determined to be 3.8~MeV/$c^2$ by calibrating the
resolution difference in the $\Xi^{-}$ mass region with the full data
sample.  The total systematic uncertainties of mass and width are
calculated to be 4.7~MeV/$c^2$ and 7.9~MeV, respectively, by summing
independent systematic sources in quadrature.

In summary, using a total of 11.0~fb$^{-1}$ of $\EE$ collision data
above the open-charm threshold collected with the BESIII detector at
the BEPCII collider, we have studied the process $\EE\ar\XXB$ based on
a single baryon tag technique. We have measured fifteen exclusive Born
cross sections and EFFs in the range from 4.009 to 4.6 GeV/$c^{2}$,
where the form factors for the process $\EE\ar\XXB$ have not been
previously measured due to limited statistics.  A fit to the dressed
cross section for $\EE\ar\XXB$ with the assumptions of a power
law dependence for continuum plus a $\psi(4230)$ or $\psi(4260)$ resonance is performed,
and no significant signal for the processes $\psi(4230)$ or $\psi(4260)\ar\XXB$ is
observed. The upper limits on the products of the electronic partial
width and the branching fractions of $\psi(4230)$ and
$\psi(4260)\ar\XXB$ are measured to be
$\Gamma_{ee}\mathcal{B}_{\psi{(4230)}} < 0.33 \times 10^{-3}$ eV and
$\Gamma_{ee}\mathcal{B}_{\psi{(4260)}} < 0.27 \times 10^{-3}$ eV at
90\% CL, which may help to understand the nature of
$\psi(4260)$~\cite{Aubert:2005rm,Close:2005iz}.  In particular, charmless
decays of the $\psi(4260)$ are expected by the hybrid
model~\cite{Close:2005iz}.  In addition, an excited $\Xi$ baryon at
$\sim$1820 MeV/$c^{2}$ is observed with a statistical significance of 6.2 $\sim$ 6.5$\sigma$ by including the systematic uncertainty, and the mass and width are measured to be $M = (1825.5
\pm 4.7 \pm 4.7)$ MeV/$c^{2}$ and $\Gamma = (17.0 \pm 15.0 \pm 7.9)$
MeV, which are consistent with the mass and width of $\Xi(1820)^{-}$
obtained from the PDG~\cite{PDG2016} within 1$\sigma$ uncertainty.
The results shed light on the structure of hyperon resonances with
strangeness $S = -2$.

\section{Acknowledgement}
\label{sec:acknowledgement}
The BESIII collaboration thanks the staff of BEPCII and the IHEP computing center for their strong support. This work is supported in part by National Key Basic Research Program of China under Contract No. 2015CB856700; 
China Postdoctoral Science Foundation under Contract No.  2018M630206;
National Natural Science Foundation of China (NSFC) under Contracts Nos. 11521505, 11625523, 11635010,  11675184, 11705209, 11735014, 11822506, 11835012, 11875115, 11905236; 
Chinese Academy of Science Focused Science Grant; National 1000 Talents Program of China;
the Chinese Academy of Sciences (CAS) Large-Scale Scientific Facility Program; Joint Large-Scale Scientific Facility Funds of the NSFC and CAS under Contracts Nos. U1532257, U1532258, U1732263, U1832207; 
CAS Key Research Program of Frontier Sciences under Contracts Nos. QYZDJ-SSW-SLH003, QYZDJ-SSW-SLH040; 
100 Talents Program of CAS; 
INPAC and Shanghai Key Laboratory for Particle Physics and Cosmology; 
ERC under Contract No. 758462; German Research Foundation DFG under Contracts Nos. Collaborative Research Center CRC 1044, FOR 2359; Istituto Nazionale di Fisica Nucleare, Italy; Koninklijke Nederlandse Akademie van Wetenschappen (KNAW) under Contract No. 530-4CDP03; Ministry of Development of Turkey under Contract No. DPT2006K-120470; National Science and Technology fund; STFC (United Kingdom); The Knut and Alice Wallenberg Foundation (Sweden) under Contract No. 2016.0157; The Royal Society, UK under Contracts Nos. DH140054, DH160214; The Swedish Research Council; U. S. Department of Energy under Contracts Nos. DE-FG02-05ER41374, DE-SC-0010118, DE-SC-0012069; University of Groningen (RuG) and the Helmholtzzentrum fuer Schwerionenforschung GmbH (GSI), Darmstadt

\end{document}